\documentclass[12pt]{article}

\usepackage{graphicx}
\usepackage{amsmath,amssymb}
\usepackage{bm}
\usepackage{graphicx, color}
\usepackage{wrapfig}
\begin{document}
\title{
\begin{flushright}
\ \\*[-80pt] 
\begin{minipage}{0.2\linewidth}
\normalsize
KUNS-2345 \\*[50pt]
\end{minipage}
\end{flushright}
{\Large \bf 
Towards Minimal $S_4$ Lepton Flavor Model
\\*[20pt]}}

\author{
\centerline{
Hajime~Ishimori, \
Tatsuo~Kobayashi, \ }  \\
\\*[20pt]
\centerline{
\begin{minipage}{\linewidth}
\begin{center}
{\it \normalsize 
Department of Physics, Kyoto University, 
Kyoto 606-8502, Japan} \\
\end{center}
\end{minipage}}
\\*[50pt]}
\vskip 2 cm
\date{\small
\centerline{ \bf Abstract}
\begin{minipage}{0.9\linewidth}
\medskip 
 We study lepton flavor models with the $S_4$ flavor symmetry.
We construct simple models with smaller numbers of flavon 
fields and free parameters, such that we have predictions 
among lepton masses and mixing angles.
The model with a $S_4$ triplet flavon is not realistic, 
but we can construct realistic models with two triplet flavons, 
or one triplet and one doublet flavons.
\end{minipage}
}

\begin{titlepage}
\maketitle
\thispagestyle{empty}
\end{titlepage}

\section{Introduction}

In particle physics, it is one of most important issues to 
understand the origin of the hierarchy among 
 quark/lepton masses and their mixing angles.
Indeed, there are many free parameters in the standard model 
including its extension with neutrino mass terms,  
and most of them are originated from the flavor sector, 
i.e. Yukawa couplings of quarks and leptons.
Recent experiments of the neutrino oscillation can 
determine neutrino mass squared  differences and 
mixing angles increasing their preciseness
~\cite{Schwetz:2008er,Fogli:2008jx,Fogli:2009zza,GonzalezGarcia:2010er,Schwetz:2011qt}.
This indicates large mixing angles, which are completely  
 different from the quark mixing ones.
In particular, the tri-bimaximal mixing is one of interesting 
An\"atze 
 in the lepton
 sector~\cite{Harrison:2002er,Harrison:2002kp,Harrison:2003aw,Harrison:2004uh}.

Non-Abelian flavor symmetries, in particular 
non-Abelian discrete symmetries, could explain such large 
mixing angles \cite{Ishimori:2010au}.
For example, by use of the $A_4$ flavor symmetry, 
the tri-bimaximal mixing of leptons has been
derived~\cite{Ma:2001dn,Ma:2002ge,Ma:2004zv,Altarelli:2005yp,Altarelli:2005yx}.
Furthermore, phenomenologically interesting aspects of 
$A_4$ flavor models have been studied~\cite{Babu:2002in}-\cite{Shimizu:2011xg}.  
Another interesting flavor symmetry is the $S_4$ symmetry~\cite{Yamanaka:1981pa,Brown:1984dk,Brown:1984mq,Ma:2005pd}.
One can realize the exact tri-bimaximal neutrino mixing 
in $S_4$ flavor
models~\cite{Lam:2008sh,Bazzocchi:2008ej,Ishimori:2008fi,Grimus:2009pg,Bazzocchi:2009pv,Bazzocchi:2009da,Meloni:2009cz}.
The $S_4$ flavor symmetry can lead other interesting aspects 
such as realistic quark mass matrices,  a grand unified theory, etc
\cite{Zhang:2006fv}-\cite{Ahn:2010nw}.

The tri-bimaximal mixing is quite interesting Ansatz at a certain
level.
For $\theta_{13}$, we have its upper bound.
It is a current experimental target to measure a finite value 
of $\theta_{13}$, and a finite value of $\theta_{13}$ would be 
measured in near future.\footnote{After this paper was completed, 
Ref.~\cite{:2011sj} appeared.}
(See also for a global fit analysis of neutrino oscillation data 
\cite{Schwetz:2011qt}, which  
suggests non-vanishing value for the mixing angle $\theta_{13}$.)
It would be straightforward to obtain non-zero $\theta_{13}$ 
by adding correction terms in the models leading to the tri-bimaximal mixing.
In this case, we may have no clear prediction on $\theta_{13}$ 
in some models, although we could keep our predictability 
on other models.
At any rate, models would become complicated.

Indeed, most of models include several flavon fields, 
whose vacuum expectation values (VEVs) break flavor symmetries.
In addition, there are many free parameters to 
derive lepton masses and mixing angles.
Thus it is important to study whether models with the minimal or smaller 
number of flavon fields can lead to realistic results 
and whether there are models with higher predictability, that is, 
that the number of free parameters is smaller than the number 
of observables such as masses and mixing angles.
Our purpose here is to study simple models with 
a small number of flavon fields and a small number of free parameters, 
such that our models have predictions on masses and mixing angles, 
e.g. their relations.

In this paper, we consider $S_4$ as the flavor symmetry 
and study simple supersymmetric model constructions with the smaller numbers of
flavons and free parameters.
When the three families correspond to a $S_4$ triplet, 
we have smaller number of free parameters.
On the other hand, when the three families correspond to a singlet 
and a doublet, couplings including the $S_4$ singlet lepton 
and $S_4$ doublet are independent of each other.
Then, we would have more free parameters.
Thus, here we concentrate on the models, in which the three families 
of both the left and right-handed leptons 
correspond to $S_4$ triplets.
Obviously the simplest model is the model with only one triplet flavon.
However, we show that such models do not lead to realistic results.
Hence, we add a $S_4$ doublet or triplet as the next simple models.
These models have seven free parameters in 
the lepton mass matrices.
Thus, they have predictions among masses and mixing angles.
Furthermore, since the neutrino mass spectrum is determined, 
the sum of neutrino masses and effective mass of 
double beta decay are also predicted. 
These predictions would be useful to search a hint 
of non-Abelian flavor symmetry $S_4$.


This paper is organized as follows. 
In section 2, we study the simple model with 
one $S_4$ triplet flavon.
Such a model is not realistic.
In section 3 we study the model with 
one triplet and one doublet flavon fields, that is, 
model III.
In section 4 we study the model 
with two triplet flavon fields, that is, model IV.
The models III and IV are realistic and have 
predictions among lepton masses and mixing angles.
In section 5, we give a comment on the model 
with the $\Delta(54)$ flavor symmetry, which is 
quite similar to the model IV.
Section 6 is devoted to the summary.


\section{Model with a triplet flavon}

The simplest model is the model with a triplet flavon.
In this section, we study such two models and show 
we can not obtain realistic results.

\subsection{Model I}
\begin{table}[h]
\begin{tabular}{|c|cc||cc|}
\hline
&$(\ell_e,\ell_\mu,\ell_\tau)$ & $(e^c,\mu^c,\tau^c)$  & $H_{u,d}$  
&$(\chi_1,\chi_2,\chi_3)$   \\ \hline
$S_{4}$ & $3$ & $3$  & $1$&  $3$   \\
\hline
\end{tabular}
\caption{Matter content and charge assignment of model I.}
\end{table}

We first consider the simplest model among all other $S_4$ models,
i.e. model I.
Each of left-handed lepton doublets and right-handed charged leptons 
are assigned to $S_4$ triplet $3$ and 
additional flavon fields $(\chi_1,\chi_2,\chi_3)$ are
 also assigned to the same triplet. 
The up and down sectors of electroweak Higgs fields are 
$S_4$ trivial singlets.
These $S_4$ representations are shown in Table 1.
In this model, the superpotential of charged leptons 
is written by
\begin{eqnarray}
\begin{split}
w_e
=&y_{1}^e(e^c\ell_e+\mu^c\ell_\mu+\tau^c\ell_\tau) H_d
\\
&+y_{2}^e
((\mu^c\ell_\tau+\tau^c\ell_\mu)\chi_1
+(e^c\ell_\tau+\tau^c\ell_e)\chi_2
+(e^c\ell_\mu+\mu^c\ell_e)\chi_3)H_d/\Lambda .
\end{split}
\end{eqnarray}
For the neutrino sector, we have
\begin{eqnarray}
\begin{split}
w_\nu
=&y_{1}^\nu(\ell_e\ell_e+\ell_\mu\ell_\mu+\ell_\tau\ell_\tau)H_uH_u/\Lambda
\\
&+y_{2}^\nu
((\ell_\mu \ell_\tau+\ell_\tau \ell_\mu)\chi_1
+(\ell_e \ell_\tau+\ell_\tau \ell_e)\chi_2
+(\ell_e\ell_\mu+\ell_\mu \ell_e)\chi_3)H_uH_u/\Lambda^2.
\end{split}
\end{eqnarray}
The VEVs of scalar fields are given by
\begin{eqnarray}
\langle H_{u,d}\rangle
=v_{u,d},
\quad
\langle \chi_{n}\rangle
=\alpha_n\Lambda.
\end{eqnarray}
We reparametrize the VEVs of $\chi_i$ for $i=1,2,3$ as
\begin{eqnarray}
\langle(\chi_{1},\chi_{2},\chi_3)\rangle
=\alpha_{1}\Lambda(1,r,r').
\end{eqnarray}
Then mass matrices become
\begin{eqnarray}
\begin{split}
M_e
=&y_1^e v_d
 \begin{pmatrix}1   & 0 & 0 \\ 
                   0    & 1  &0    \\
                   0  & 0  & 1    \\
 \end{pmatrix} 
+y_2^e\alpha_1 v_d
\begin{pmatrix}
0&r'&r\\
r'&0&1\\
r&1&0
\end{pmatrix},
 \\
M_\nu
=&y_1^\nu \frac{v_u^2}{\Lambda}
 \begin{pmatrix}1   & 0 & 0 \\ 
                   0    & 1  &0    \\
                   0  & 0  & 1    \\
 \end{pmatrix} 
+y_2^\nu\alpha_1 \frac{v_u^2}{\Lambda}
\begin{pmatrix}
0&r'&r\\
r'&0&1\\
r&1&0
\end{pmatrix}.
\end{split}
\end{eqnarray}
The off-diagonal elements of 
the neutrino mass matrix are the same as the ones of charged 
leptons.
Indeed, we can rewrite the mass matrix of 
neutrinos as
\begin{eqnarray}
M_\nu
=(y_1^\nu -\frac{y_1^ey_2^\nu}{y_2^e})\frac{v_u^2}{\Lambda}
 \begin{pmatrix}1   & 0 & 0 \\ 
                   0    & 1  &0    \\
                   0  & 0  & 1    \\
 \end{pmatrix} 
~+~\frac{y_2^\nu v_u^2}{y_2^ev_d\Lambda} M_e .
\end{eqnarray}
Then, one can not realize large mixing angles.
With a non-vanishing CP-phase, the mixing matrix does not need 
to be trivial, but we cannot obtain large mixing 
angles indicated by experiments of neutrino oscillation. 
We could introduce a $Z_N$ symmetry such that it 
allows either $y_2^e$ or $y_2^\nu$.
In this case, one could not realize realistic mass eigenvalues.

\subsection{Model II}
\begin{table}[h]
\begin{tabular}{|c|cc||ccc|}
\hline
&$(\ell_e,\ell_\mu,\ell_\tau)$ & $(e^c,\mu^c,\tau^c)$  & $H_{u,d}$  
&$\chi_1$&$(\chi_2,\chi_3,\chi_4)$   \\ \hline
$S_{4}$ & $3$ & $3'$  & $1$&$1'$&  $3$   \\
\hline
\end{tabular}
\caption{Matter content and charge assignment of model II.}
\end{table}

Here, we discuss another model with a $S_4$ triplet flavon 
as model II.
As indicated by the model I, 
if the off-diagonal elements of 
charged leptons and neutrinos are the same, 
realistic lepton mixing cannot be obtained. 
Another candidate for the simplest model is given by 
changing the $S_4$ charge assignment. 
Lepton doublets are assigned to 
$3$ while right handed-charged leptons 
are assigned to $3'$ of $S_4$. 
In addition, we consider $S_4$ singlet flavon $\chi_1$ with 
the charge $1'$ and triplet flavon $(\chi_2,\chi_3,\chi_4)$ 
with $3'$. 
These $S_4$ representations are shown in Table 2.
Then the flavor symmetric superpotential becomes
\begin{eqnarray}
\begin{split}
w_e
=&y_{1}^e(e^c\ell_e+\mu^c\ell_\mu+\tau^c\ell_\tau)\chi_1 H_d
\\
&+y_{2}^e
((\tau^c\ell_\mu-\mu^c\ell_\tau)\chi_2
+(e^c\ell_\tau-\tau^c\ell_e)\chi_3
+(\mu^c\ell_e-e^c\ell_\mu)\chi_4)H_d/\Lambda,
\end{split}
\end{eqnarray}
for charged leptons and 
\begin{eqnarray}
\begin{split}
w_\nu
=&y_{1}^\nu(\ell_e\ell_e+\ell_\mu\ell_\mu+\ell_\tau\ell_\tau)H_uH_u/\Lambda
\\
&+y_{2}^\nu
((\ell_\mu \ell_\tau+\ell_\tau \ell_\mu)\chi_2
+(\ell_e \ell_\tau+\ell_\tau \ell_e)\chi_3
+(\ell_e\ell_\mu+\ell_\mu \ell_e)\chi_4)H_uH_u/\Lambda^2,
\end{split}
\end{eqnarray}
for neutrinos. 
Vacuum expectation values are given by
\begin{eqnarray}
\langle H_{u,d}\rangle
=v_{u,d},
\quad
\langle \chi_{1}\rangle
=\alpha_1\Lambda,
\quad
\langle(\chi_{2},\chi_{3},\chi_4)\rangle
=\alpha_{2}\Lambda(1,r,r').
\end{eqnarray}
Then, mass matrices are obtained 
\begin{eqnarray}
\begin{split}
M_e
=&y_1^e \alpha_1 v_d
 \begin{pmatrix}1   & 0 & 0 \\ 
                   0    & 1  &0    \\
                   0  & 0  & 1    \\
 \end{pmatrix} 
+y_2^e\alpha_2 v_d
\begin{pmatrix}
0&-r'&r\\
r'&0&-1\\
-r&1&0
\end{pmatrix},
 \\
M_\nu
=&y_1^\nu \frac{v_u^2}{\Lambda}
 \begin{pmatrix}1   & 0 & 0 \\ 
                   0    & 1  &0    \\
                   0  & 0  & 1    \\
 \end{pmatrix} 
+y_2^\nu\alpha_2 \frac{v_u^2}{\Lambda}
\begin{pmatrix}
0&r'&r\\
r'&0&1\\
r&1&0
\end{pmatrix}.
\end{split}
\end{eqnarray}
Let us consider the limit $m_e=0$.
Then, the determinant of the charged lepton mass matrix 
can be vanishing when $y_1^e\alpha_1/y_2^e\alpha_2=0,\pm\sqrt{-1-r^2-r'^2}$. 
For the first case, we have $m_\mu=m_\tau$ and 
for the other cases, $2m_\mu=m_\tau$. 
Then this model cannot lead the realistic mass spectrum of charged 
leptons.

\section{Model III}

In the previous section, it was shown that the models
with a $S_4$ triplet flavon does not lead to 
realistic results.
Thus, the next step is to add another flavon 
with non-trivial $S_4$ representations.
In this section we add a $S_4$ doublet flavon, 
and in the next section we add a $S_4$ triplet flavon.

\subsection{Mass matrices}

\begin{table}[h]
\begin{tabular}{|c|cc||ccccc|}
\hline
&$(\ell_e,\ell_\mu,\ell_\tau)$ & $(e^c,\mu^c,\tau^c)$  & $H_{u,d}$  
&$ \chi_0$&$ \chi_1$ &$(\chi_2,\chi_3)$&$(\chi_4,\chi_5,\chi_6)$   \\ \hline
$S_{4}$ & $3$ & $3'$  & $1$& $1$& $1$&$2$&  $3'$   \\
$Z_{3}$ & $1$ & $1$ & $0$& $0$ & $1$&$1$&  $1$   \\
\hline
\end{tabular}
\caption{Matter content and charge assignment of model III.}
\end{table}

In this model, we add one $S_4$ doublet. 
Then, we can fit experimental values. 
The charge assignments for leptons 
and flavons are summarized in Table 3. 
The field $\chi_0$ is added to realize a proper pattern of 
the vacuum alignment as will be discussed.
To make stronger prediction, we assume 
there is no mixing from the neutrino sector 
which is realized in the charge assignment with 
the $Z_3$ symmetry. 
The superpotential of charged leptons is
\begin{eqnarray}
\begin{split}
w_e
=&y_{1}^e(e^c\ell_e+\mu^c\ell_\mu+\tau^c\ell_\tau)\chi_1 H_d/\Lambda
\\
&+y_{2}^e(\frac1{\sqrt2}(\mu^c\ell_\mu-\tau^c\ell_\tau)\chi_3
+\frac1{\sqrt6}(2e^c\ell_e-\mu^c\ell_\mu-\tau^c\ell_\tau)\chi_2) H_d/\Lambda
\\
&+y_{3}^e
((\tau^c\ell_\mu+\mu^c\ell_\tau)\chi_3
+(e^c\ell_\tau+\tau^c\ell_e)\chi_4
+(\mu^c\ell_e+e^c\ell_\mu)\chi_5)H_d/\Lambda .
\end{split}
\end{eqnarray}
Similarly, the superpotential of neutrinos is obtained 
\begin{eqnarray}
\begin{split}
w_\nu
=&y_{1}^\nu(\ell_e\ell_e+\ell_\mu\ell_\mu+\ell_\tau\ell_\tau)\chi_1H_uH_u/\Lambda^2
\\
&+y_{2}^\nu(\frac{1}{\sqrt2}
(\ell_\mu\ell_\mu-\ell_\tau\ell_\tau)\chi_2
+\frac1{\sqrt6}
(-2\ell_e\ell_e+\ell_\mu\ell_\mu+\ell_\tau\ell_\tau)\chi_3)H_uH_u/\Lambda.
\end{split}
\end{eqnarray}
VEVs are denoted by
\begin{eqnarray}
\langle H_{u,d}\rangle
=v_{u,d},
\quad
\langle \chi_{n}\rangle
=\alpha_n\Lambda.
\end{eqnarray}
The vacuum alignment is assumed to be
\begin{eqnarray}
\langle(\chi_{2},\chi_3)\rangle
=\alpha_{2}\Lambda(1,r),
\quad
\langle(\chi_{4},\chi_{5},\chi_6)\rangle
=\alpha_{4}\Lambda(1,r',r').
\end{eqnarray}
Then, mass matrices are written 
\begin{eqnarray}
\begin{split}
M_e
=&y_1^e \alpha_1 v_d
 \begin{pmatrix}1   & 0 & 0 \\ 
                   0    & 1  &0    \\
                   0  & 0  & 1    \\
 \end{pmatrix} 
+y_2^e \alpha_2 v_d
 \begin{pmatrix}\frac{2}{\sqrt6}   & 0 & 0 \\ 
                   0    & \frac{r}{\sqrt2}-\frac{1}{\sqrt6}  &0    \\
                   0  & 0  & -\frac{r}{\sqrt2}-\frac{1}{\sqrt6}   \\
 \end{pmatrix} 
+y_3^e\alpha_4 v_d
\begin{pmatrix}
0&r'&r'\\
r'&0&1\\
r'&1&0
\end{pmatrix},
 \\
M_\nu
=&y_1^\nu\alpha_1 \frac{v_u^2}{\Lambda}
 \begin{pmatrix}1   & 0 & 0 \\ 
                   0    & 1  &0    \\
                   0  & 0  & 1    \\
 \end{pmatrix} 
+y_2^\nu\alpha_2 \frac{v_u^2}{\Lambda}
 \begin{pmatrix}-\frac{2r}{\sqrt6}   & 0 & 0 \\ 
                   0    & \frac{1}{\sqrt2}+\frac{r}{\sqrt6}  &0    \\
                   0  & 0  & -\frac{1}{\sqrt2}+\frac{r}{\sqrt6}   \\
 \end{pmatrix} .
\end{split}
\end{eqnarray}
Suppose that the charged lepton mass matrix is diagonalized by 
$U_e$.
The neutrino mass matrix in the basis 
of the diagonal charged lepton mass matrix is written 
\begin{eqnarray}
\begin{split}
U_e^TM_\nu U_e
=&\frac{y_3^\nu}{y_3^e}\frac{v_u^2}{v_d\Lambda}
U_e^T U_e
\begin{pmatrix}
m_e&0&0\\
0&m_\mu&0\\
0&0&m_\tau
\end{pmatrix}
+(y_1^\nu -\frac{y_3^\nu y_1^e}{y_3^e})\frac{v_u^2}{\Lambda}
U_e^T
 \begin{pmatrix}1   & 0 & 0 \\ 
                   0    & 1  &0    \\
                   0  & 0  & 1    \\
 \end{pmatrix} U_e
\\
&+(y_2^\nu-\frac{y_2^ey_3^\nu}{y_3^e})\alpha_1 \frac{v_u^2}{\Lambda}
U_e^T 
\begin{pmatrix}-\frac{2r}{\sqrt6}   & 0 & 0 \\ 
                   0    & \frac{1}{\sqrt2}+\frac{r}{\sqrt6}  &0    \\
                   0  & 0  & -\frac{1}{\sqrt2}+\frac{r}{\sqrt6}   \\
 \end{pmatrix} 
U_e.
\end{split}
\end{eqnarray}
In this case, by introducing the $S_4$ doublet flavon,
we have more free parameters in $M_e$ 
compared with the mass matrices in the previous section.
Then we can obtain realistic values of charged lepton masses and 
two large mixing angles.

There are five parameters for the charged lepton mass matrix
so that they can be fixed by giving 
two mixing angles of leptons and charged lepton masses.
The other mixing angle is determined. 
After fixing the parameters, the neutrino mass matrix 
has two degrees of freedom which can be 
determined by mass squared differences of 
neutrinos. 
For instance, when we give $\sin^2\theta_{12}^\text{MNS}=1/3$, 
$\sin^2\theta_{23}^\text{MNS}=1/2$, 
we obtain $\theta_{13}^\text{MNS}$ and the mass spectrum 
of neutrinos. 
Such angles as well as charged lepton mass ratios are realized when 
$r'=0.567$, $y_1^e/y_3^e\alpha_3=0.857$, 
$y_2^e/y_3^e\alpha_3=-0.225$, and $r=-2.83$, 
then we obtain 
\begin{eqnarray}
M_e\approx 
y_3\alpha_4 v_d
\begin{pmatrix}
0.338&0.567&0.567\\
0.567&0.957&1\\
0.567&1&1.27
\end{pmatrix},
\quad
U_e^T\approx 
\begin{pmatrix}
0.813&-0.575&0.0895\\
-0.460&-0.541&0.704\\
-0.357&-0.614&-0.704
\end{pmatrix}.
\end{eqnarray}
The predicted value for  $\sin\theta_{13}^\text{MNS}$ is $\sin\theta_{13}^\text{MNS}\approx 0.0895$. 
For neutrino masses, we have
\begin{eqnarray}
\begin{split}
&m_{\nu_1}
=y_2^\nu\alpha_1\frac{v_u^2}{\Lambda}
(\frac{y_1^\nu\alpha_1}{y_2^\nu\alpha_2}-\frac{2r}{\sqrt6}),
\quad
m_{\nu_2}
=y_2^\nu\alpha_1\frac{v_u^2}{\Lambda}
(\frac{y_1^\nu\alpha_1}{y_2^\nu\alpha_2}+\frac{1}{\sqrt2}+\frac{r}{\sqrt6}),
\\
&m_{\nu_3}
=y_2^\nu\alpha_1\frac{v_u^2}{\Lambda}
(\frac{y_1^\nu\alpha_1}{y_2^\nu\alpha_2}-\frac{1}{\sqrt2}+\frac{r}{\sqrt6}).
\end{split}
\end{eqnarray}
Mass squared differences of atmospheric and solar are 
obtained by assuming normal hierarchy. 
Writing $\text{Arg}(\frac{y_1^\nu\alpha_1}{y_2^\nu\alpha_2})
=a$, we have
\begin{eqnarray}
\frac{y_1^\nu\alpha_1}{y_2^\nu\alpha_2}
\approx\frac{1.93e^{2ia}}{1+e^{2ia}},
\quad
|y_2^\nu|\alpha_1\frac{v_u^2}{\Lambda}
\approx
0.0201 [\text{eV}].
\end{eqnarray}
>From them, the lowest value of the sum of neutrino mass becomes 
$\sum m_{\nu_i}\approx 0.113$eV and of the effective mass of 
double beta decay is $|m_{ee}|\approx 8.05$meV.

\begin{figure}
\begin{center}
\includegraphics[width=7.5cm]{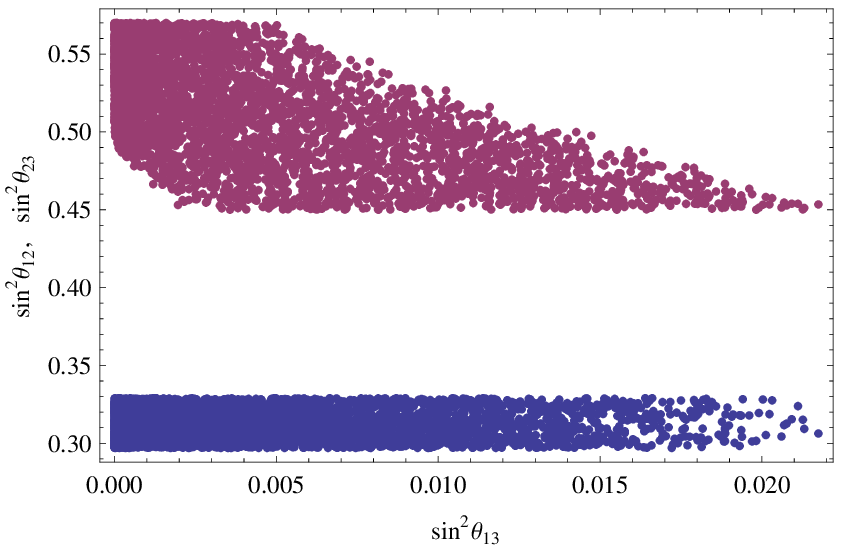}
\quad
\includegraphics[width=7.5cm]{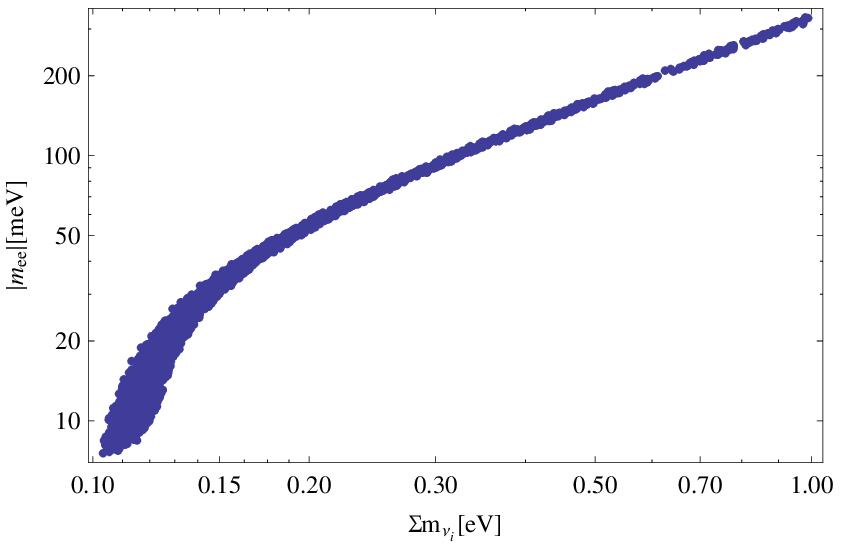}
\caption{Predicted values of mixing angles (left) 
and neutrino masses (right). For the left figure, 
blue dots (lower dots) denote 
$\sin^2\theta_{13}$--$\sin^2\theta_{12}$ plane and 
red dots (upper dots) denote 
$\sin^2\theta_{13}$--$\sin^2\theta_{23}$ plane. }
\end{center}
\end{figure}

In numerical calculation, we input $\theta_{12}^\text{MNS}$ and 
$\theta_{23}^\text{MNS}$ within 1$\sigma$ range of \cite{Schwetz:2011qt} 
for the case of normal mass hierarchy:
\begin{eqnarray}
&&\Delta m_{31}^2=(2.36- 2.54)\times 10^{-3} 
{\rm eV}^2 \ ,
\quad \Delta m_{21}^2= (7.41- 7.79) \times 10^{-5} {\rm eV}^2  \ , 
\nonumber \\
&& \sin^2 \theta_{12}=0.297- 0.329 \ ,
\quad  \sin^2 \theta_{23}=0.45 -0.57 .
\end{eqnarray} 
With these values for $\theta_{12}^\text{MNS}$ and 
$\theta_{23}^\text{MNS}$, we obtain $\theta_{13}^\text{MNS}$, 
$\sum m_{\nu_i}$, and $|m_{ee}|$ for each value, as 
indicated by Fig. 1. 
Important predictions of this model are the correlation of 
$\sin^2\theta_{13}$--$\sin^2\theta_{23}$ with narrow band 
and allowed region of $\sum m_{\nu_i}$--$|m_{ee}|$ plane.
Considering improvement of experiments in future, 
prediction would be stronger, depending on the parameter 
region of input values. Precise measurement of 
mixing angles can test the model in near future. 
For neutrino masses, they can be also improved 
by the precise values of input mixing angles.

\subsection{Potential analysis}

The superpotential including only the flavon fields is
obtained as 
\begin{eqnarray}
\begin{split}
w_s=&\kappa\chi_0^2
+\lambda_1\chi_1^3
+\lambda_2\chi_1(\chi_2^2+\chi_3^2)
+\lambda_3\chi_1(\chi_4^2+\chi_5^2+\chi_6^2)
\\&
+\lambda_4(3\chi_2^2\chi_3-\chi_3^3)
+\lambda_5(\frac1{\sqrt2}\chi_2(\chi_5^2-\chi_6^2)
+\frac1{\sqrt6}\chi_3(-2\chi_4^2+\chi_5^2+\chi_6^2))
\\
&+\eta_1\chi_0^4
+6\eta_2\chi_0\chi_4\chi_5\chi_6.
\end{split}
\end{eqnarray}
The conditions of potential minimum read
\begin{eqnarray}
\begin{split}
2\kappa\chi_0
+4\eta_1\chi_0^3
+6\eta_2\chi_4\chi_5\chi_6
&=0,
\\
3\lambda_1\chi_1^2
+\lambda_2(\chi_2^2+\chi_3^2)
+\lambda_3(\chi_4^2+\chi_5^2+\chi_6^2)
&=0,
\\
2\kappa_2\chi_2+
\lambda_{12}(\chi_3^3-3\chi_3\chi_4^2)
+6\lambda_{14}\chi_5\chi_6\chi_7&=0,
\\
2\lambda_2\chi_1\chi_2
+6\lambda_4\chi_2\chi_3
+\frac{1}{\sqrt2}\lambda_5(\chi_5^2-\chi_6^2)
&=0,
\\
2\lambda_2\chi_1\chi_3
+3\lambda_4(\chi_2^2-\chi_3^2)
+\frac{1}{\sqrt6}\lambda_5(-2\chi_4^2+\chi_5^2+\chi_6^2)
&=0,
\\
2\lambda_3\chi_1\chi_4
-\frac{4}{\sqrt6}\lambda_5\chi_3\chi_4
+6\eta_2\chi_0\chi_5\chi_6
&=0,
\\
2\lambda_3\chi_1\chi_5
+2\lambda_5\chi_2\chi_5
+\frac{2}{\sqrt6}\lambda_5\chi_3\chi_5
+6\eta_2\chi_0\chi_4\chi_6
&=0,
\\
2\lambda_3\chi_1\chi_6
-2\lambda_5\chi_2\chi_6
+\frac{2}{\sqrt6}\lambda_5\chi_3\chi_6
+6\eta_2\chi_0\chi_4\chi_5
&=0.
\end{split}
\end{eqnarray}
Since there are more than six parameters, 
it is easy to obtain independent values for 
all VEVs. 
Note that the alignment with strict relation $\chi_5=\chi_6$ 
leads $\lambda_5=0$ from the last three equations. 
Taking this, it automatically makes $\chi_4=\chi_5=\chi_6$ 
if they have non-vanishing VEVs. 
Then the vacuum alignment of the model 
must be interpreted as 
$\chi_5\approx \chi_6$. 
Choosing some parameter region of the above superpotential, 
this relation  holds so that the same result can be obtained.

\section{Model IV}

\subsection{Mass matrices}

\begin{table}[h]
\begin{tabular}{|c|cc||ccccc|}
\hline
&$(\ell_e,\ell_\mu,\ell_\tau)$ & $(e^c,\mu^c,\tau^c)$  & $H_{u,d}$  
&$\chi_1$&$\chi_2$
&$(\chi_3,\chi_4,\chi_5)$&$(\chi_6,\chi_7,\chi_8)$   \\ \hline
$S_{4}$ & $3$ & $3$  & $1$& $1$& $1$&$3$&  $3$   \\
$Z_{3}$ & $1$ & $0$  & $0$&$2$
&$1$&$2$&  $1$   \\
\hline
\end{tabular}
\caption{Matter content and charge assignment of model IV.}
\end{table}

Here, we study the model with two $S_4$ triplet flavons, 
that is, model IV.
In this model, each of charged lepton sector 
and neutrino sector couples to (different) $S_4$ triplet flavon. 
The pattern of mass matrices is the same between 
the charged leptons and neutrinos.  
Then the maximal number of parameters is equal to four in each sector. 
Considering a proper pattern of the vacuum alignment, 
the model has some prediction. 
The $S_4$ representations and $Z_3$ charges are shown in Table 4.
Now let us study the prediction of this model. 
The superpotential of charged leptons is written by 
\begin{eqnarray}\label{eq:model4-W-L}
\begin{split}
w_e
=&y_{1}^e(e^c\ell_e+\mu^c\ell_\mu+\tau^c\ell_\tau)\chi_1 H_d/\Lambda
\\
&+y_{2}^e
((\tau^c\ell_\mu+\mu^c\ell_\tau)\chi_3
+(e^c\ell_\tau+\tau^c\ell_e)\chi_4
+(\mu^c\ell_e+e^c\ell_\mu)\chi_5)H_d/\Lambda .
\end{split}
\end{eqnarray}
The superpotential including neutrinos is written 
\begin{eqnarray}\label{eq:model4-W-nu}
\begin{split}
w_\nu
=&y_{1}^\nu(\ell_e\ell_e+\ell_\mu\ell_\mu+\ell_\tau\ell_\tau)\chi_2H_uH_u/\Lambda^2
\\
&+y_{2}^\nu
((\ell_\mu \ell_\tau+\ell_\tau \ell_\mu)\chi_6
+(\ell_e \ell_\tau+\ell_\tau \ell_e)\chi_7
+(\ell_e\ell_\mu+\ell_\mu \ell_e)\chi_8)H_uH_u/\Lambda^2.
\end{split}
\end{eqnarray}
The vacuum alignment is assumed to be
\begin{eqnarray}\label{eq:model-4-vacuum}
\langle(\chi_{3},\chi_4,\chi_5)\rangle
=\alpha_{3}\Lambda(1,1,r),
\quad
\langle(\chi_{6},\chi_{7},\chi_8)\rangle
=\alpha_{7}\Lambda(1,r',r'').
\end{eqnarray}
Then the mass matrices are given
\begin{eqnarray}
\begin{split}
M_e
=&y_1^e  v_d
 \begin{pmatrix}1   & 0 & 0 \\ 
                   0    & 1  &0    \\
                   0  & 0  & 1    \\
 \end{pmatrix} 
+y_2^e\alpha_3 v_d
\begin{pmatrix}
0&r&1\\
r&0&1\\
1&1&0
\end{pmatrix},
 \\
M_\nu
=&y_1^\nu \frac{v_u^2}{\Lambda}
 \begin{pmatrix}1   & 0 & 0 \\ 
                   0    & 1  &0    \\
                   0  & 0  & 1    \\
 \end{pmatrix} 
+y_2^\nu\alpha_7 \frac{v_u^2}{\Lambda}
\begin{pmatrix}
0&r''&r'\\
r''&0&1\\
r'&1&0
\end{pmatrix}.
\end{split}
\end{eqnarray}
For the charged leptons, 
there remains the $e-\mu$ symmetry so that
\begin{eqnarray}
\begin{split}
U_{12}
=\begin{pmatrix}
1/\sqrt2 &1/\sqrt2&0\\
-1/\sqrt2 &1/\sqrt2&0\\
0&0&1\\
\end{pmatrix},
\quad
&U_{12}^{\dagger}
M_{e} U_{12}
=y_2^e\alpha_3 v_d
\begin{pmatrix}
a-r&0&0\\
0&a+r&\sqrt 2\\
0&\sqrt 2&a
\end{pmatrix},
\end{split}
\end{eqnarray}
where $a=y_1^e\alpha_1/y_2^e\alpha_2$. 
Then the mass matrix $M_e$ can be diagonalized by
\begin{eqnarray}
\begin{split}
&U_{23}
=\begin{pmatrix}
1  &0&0\\
0&\cos\theta_{23} &-\sin\theta_{23} \\
0&\sin\theta_{23}&\cos\theta_{23}
\end{pmatrix},
\quad
\tan \theta_{23}
=\frac{-r+\sqrt{8+r^2}}{2\sqrt2},
\\
&U_{23}^{e\dagger}
U_{12}^{e\dagger}
M_{e} U^e_{12}U^e_{23}
=y_2^e\alpha_2 v_d
\begin{pmatrix}
a-r &0&0\\
0&\frac12(2a+r+\sqrt{8+r^2})&0\\
0&0&\frac12(2a+r-\sqrt{8+r^2}) 
\end{pmatrix}.
\end{split}
\end{eqnarray}
We use the notation of 
$U_{12}^{e\dagger}U_{23}^{e\dagger}
M_{e} U^e_{23}U^e_{12}=\text{diag}(m_1^e,m_2^e,m_3^e)$.
The mass matrix $M_e$ has three parameters and  
they can be fixed by masses of charged leptons. 

The parameters of the neutrino mass matrix 
are independent of the ones of charged leptons 
so that four parameters  remain. 
Using them, we need to fit two mass scales of 
neutrino oscillations and three mixing angles. 
Similar to the previous section, 
giving two mixing angles of the MNS matrix, 
the other angle and neutrino mass spectrum can be predicted. 
To fit the parameters, we write
\begin{eqnarray}
\label{numass}
M_\nu
=U_\nu
\begin{pmatrix}
m^\nu_{1}&0&0\\
0&m^\nu_{2}&0\\
0&0&m^\nu_{3}
\end{pmatrix}
U_\nu^\dagger.
\end{eqnarray}
As an example, 
assuming $m_1^e=m_e$, $m_2^e=m_\mu$, $m_3^e=m_\tau$, 
we have
\begin{eqnarray}
a\approx -1\pm\frac{9m_\mu}{4m_\tau},
\quad
r\approx 1\mp\frac{9m_\mu}{4m_\tau}.
\end{eqnarray}
For neutrino masses, we assume
$m_1^\nu=m_{\nu_1}$, $m_2^\nu=m_{\nu_2}$, $m_3^\nu=m_{\nu_3}$, 
then we have 
\begin{eqnarray}
U_\nu
=U_{12}^{e}U_{23}^{e}
U_\text{MNS}.
\end{eqnarray}
Inserting this matrix to Eq. (\ref{numass}), 
assuming $\sin^2\theta_{12}^\text{MNS}=1/3$, 
$\sin^2\theta_{23}^\text{MNS}=1/2$, 
we obtain $\sin\theta_{13}^\text{MNS}=0.1487,~0.1130$. 
For neutrino masses, there appears the following condition
\begin{eqnarray}
\frac{m_{\nu_3}}{m_{\nu_1}}
\approx
0.951(0.0518
+\frac{m_{\nu_2}}{m_{\nu_1}}),
\end{eqnarray}
for $\sin\theta_{13}^\text{MNS}=0.1487$ and 
\begin{eqnarray}
\frac{m_{\nu_3}}{m_{\nu_1}}
\approx
0.973(0.0281
+\frac{m_{\nu_2}}{m_{\nu_1}}),
\end{eqnarray}
for  $\sin\theta_{13}^\text{MNS}= 0.1130$. 
To be consistent with experiments, 
only inverted hierarchy is allowed. 
When Majorana phase is vanishing, 
the mass spectrum for $\sin\theta_{13}=0.1487$ is 
obtained 
\begin{eqnarray}
m_{\nu_1}
\approx 0.01374[\text{eV}],
\quad
m_{\nu_2}
\approx -0.01378[\text{eV}],
\quad
m_{\nu_3}
\approx -0.01242[\text{eV}].
\end{eqnarray}
The sum of neutrino masses is 
$0.341$eV and the effective mass of 
double beta decay is $35.8$meV. 
Parameters are chosen as 
$r'\approx -0.9418$, $r''\approx -0.8767$, 
$y_1^\nu/y_2^\nu\alpha_7\approx -0.4364$. 
For $\sin\theta_{13}=0.1130$, we have
\begin{eqnarray}
m_{\nu_1}
\approx 0.02429[\text{eV}],
\quad
m_{\nu_2}
\approx -0.02433[\text{eV}],
\quad
m_{\nu_3}
\approx -0.02300[\text{eV}].
\end{eqnarray}
The sum is $0.460$eV and the mass of 
double beta decay is $49.3$meV. 
Parameters are set as 
$r'\approx -0.9741$, $r''\approx -0.9311$, 
$y_1^\nu/y_2^\nu\alpha_7\approx -0.4653$.

Similar to the previous section, we input $\theta_{12}^\text{MNS}$ and 
$\theta_{23}^\text{MNS}$ within 1$\sigma$ range of \cite{Schwetz:2011qt}. 
In the above case, we can only have inverted hierarchy for 
neutrino masses, but in general case, normal hierarchy is also allowed. 
For inverted mass hierarchy, different parameter space is favoured:
\begin{eqnarray}
&&\Delta m_{31}^2=-(2.25- 2.44)\times 10^{-3} 
{\rm eV}^2 \ ,
\quad \Delta m_{21}^2= (7.41- 7.79) \times 10^{-5} {\rm eV}^2  \ , 
\nonumber \\
&& \sin^2 \theta_{12}=0.297- 0.329 \ ,
\quad  \sin^2 \theta_{23}=0.46 -0.58 .
\end{eqnarray} 
Giving $\theta_{12}^\text{MNS}$ and 
$\theta_{23}^\text{MNS}$, we can get $\theta_{13}^\text{MNS}$, 
$\sum m_{\nu_i}$, and $|m_{ee}|$ for each value, 
shown in Fig. 2. 
For the case of inverted mass hierarchy, 
the allowed region is narrow and our prediction is strong.
Lower bounds of $\sum m_{\nu_i}$ and $|m_{ee}|$ 
would be reached by next generation experiments.
This model can be also tested by precise measurement of mixing angles. 
For neutrino masses, $\sum m_{\nu_i}$ and $|m_{ee}|$ 
are expected with larger values compared to model III. 
With next-generation experiments of double beta decay 
and neutrino oscillation, we can have a hint of this model.

\begin{figure}
\begin{center}
\includegraphics[width=7.5cm]{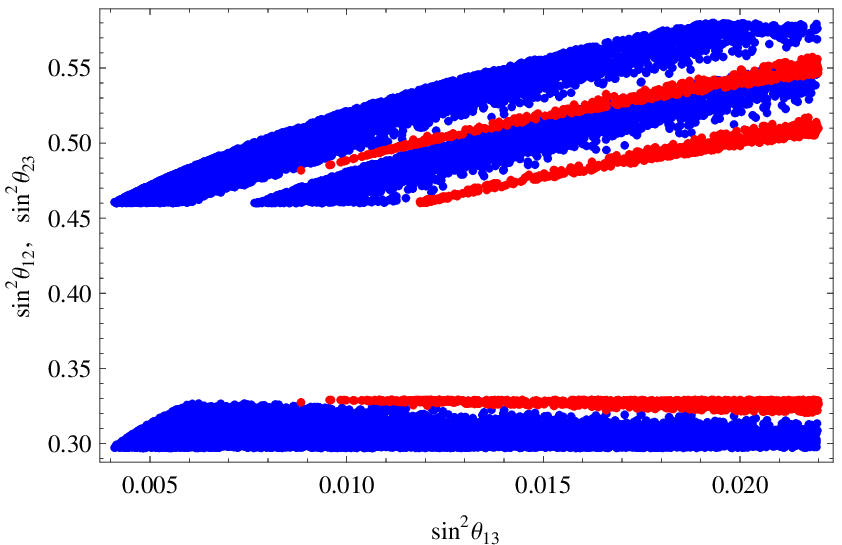}
\quad
\includegraphics[width=7.5cm]{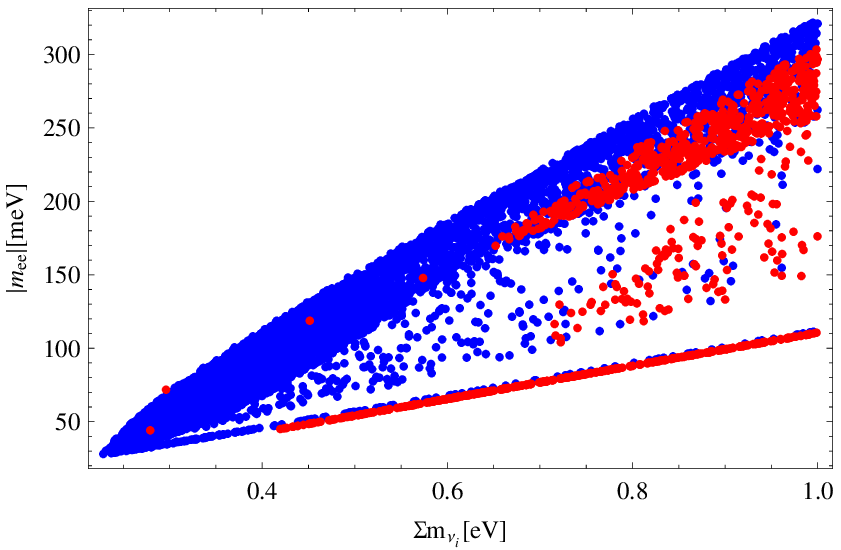}
\caption{Predicted values of mixing angles (left) 
and neutrino masses (right). For both figures, blue (dark gray) 
dots indicate normal mass hierarchy of neutrinos and 
red (light gray) dots indicate inverted mass hierarchy. }
\end{center}
\end{figure}

\subsection{Potential analysis}
The superpotential including only the flavon fields is written as 
\begin{eqnarray}
\begin{split}
w_s=&\kappa_1\chi_1\chi_2
+\kappa_2(\chi_3\chi_6+\chi_4\chi_7+\chi_5\chi_8)
+\lambda_1\chi_1^3
+\lambda_2\chi_2^3
+6\lambda_3\chi_3\chi_4\chi_5
+6\lambda_4\chi_6\chi_7\chi_8
\\
&+\lambda_5\chi_1(\chi_3^2+\chi_4^2+\chi_5^2)
+\lambda_6\chi_2(\chi_6^2+\chi_7^2+\chi_8^2)
+\eta_1(\chi_3^2+\chi_4^2+\chi_5^2)(\chi_6^2+\chi_7^2+\chi_8^2)
\\
&+\eta_1'(\frac12(\chi_4^2-\chi_5^2)( \chi_7^2-\chi_8^2)
+\frac16(-2\chi_3^2+\chi_4^2+\chi_5^2)(-2\chi_6^2+\chi_7^2+\chi_8^2))
\\
&+4\eta_1''(\chi_4\chi_5\chi_7\chi_8
+\chi_3\chi_5\chi_6\chi_8
+\chi_3\chi_4\chi_6\chi_7)+\cdots,
\end{split}
\end{eqnarray}
where we omit other fourth couplings which 
have negative mass dimension. 
Assuming $\eta_1''$ is larger than other 
negative dimensional operators, 
the condition of potential minimum becomes
\begin{eqnarray}
\begin{split}
\kappa_1\chi_2
+3\eta_1\chi_1^2
+\eta_5(\chi_3^2+\chi_4^2+\chi_5^2)
&=0,
\\
\kappa_1\chi_1
+3\eta_2\chi_2^2
+\eta_6(\chi_6^2+\chi_7^2+\chi_8^2)
&=0,
\\
\kappa_2\chi_6
+6\eta_3\chi_4\chi_5
+2\eta_5\chi_1\chi_3
+4\eta_1''(\chi_5\chi_6\chi_8+\chi_4\chi_6\chi_7)
&=0,
\\
\kappa_2\chi_7
+6\eta_3\chi_3\chi_5
+2\eta_5\chi_1\chi_4
+4\eta_1''(\chi_5\chi_7\chi_8+\chi_3\chi_6\chi_7)
&=0,
\\
\kappa_2\chi_8
+6\eta_3\chi_3\chi_4
+2\eta_5\chi_1\chi_5
+4\eta_1''(\chi_4\chi_7\chi_8+\chi_3\chi_6\chi_8)
&=0,
\\
\kappa_2\chi_3
+6\eta_4\chi_7\chi_8
+2\eta_6\chi_2\chi_6
+4\eta_1''(\chi_3\chi_5\chi_8+\chi_3\chi_4\chi_7)
&=0,
\\
\kappa_2\chi_4
+6\eta_4\chi_6\chi_8
+2\eta_6\chi_2\chi_7
+4\eta_1''(\chi_4\chi_5\chi_8+\chi_3\chi_4\chi_6)
&=0,
\\
\kappa_2\chi_5
+6\eta_4\chi_6\chi_7
+2\eta_6\chi_2\chi_8
+4\eta_1''(\chi_4\chi_5\chi_7+\chi_3\chi_5\chi_6)
&=0.
\end{split}
\end{eqnarray}
There are many parameters enough to take 
independent values for each VEV. 
To realize the alignment $\chi_3=\chi_4$ with $\chi_6\not=\chi_7$, 
we need a condition $\kappa_2+4\eta_1''\chi_5\chi_8=0$.

\section{$\Delta(54)$ model and its stringy origin}

Here, we give comments on $\Delta(54)$ models.
The $\Delta(54)$ symmetry has a structure similar to 
$S_4$.
Indeed, several interesting flavor models have been constructed 
\cite{Ishimori:2008uc,Ishimori:2009ew,King:2009ap,Escobar:2011mq}.
Furthermore, the $\Delta(54)$ flavor symmetry 
as well as $D_4$ and $\Delta(27)$ can be realized 
within the framework of heterotic string models on orbifolds 
\cite{Kobayashi:2004ya,Kobayashi:2006wq,Ko:2007dz}
and magnetized/intersecting D-brane models \cite{Abe:2009vi,Abe:2010ii}.
\footnote{See also \cite{Abe:2010iv}.}
In particular, only triplets as well as a trivial singlet 
appear as fundamental modes in heterotic orbifold models 
\cite{Kobayashi:2006wq}. From this viewpoint, 
the model in section 4 is quite interesting.

We assign $\Delta(54)$ representations and $Z_3$ charges to 
the leptons and flavons as shown in Table 5.
Those are the same  
as the model in section 4 except replacing $S_4$ by $\Delta(54)$.
The $Z_3$ charge assignment is also the same.
This $Z_3$ symmetry plays a role such that 
different triplet flavon VEVs appear in the mass matrices of 
the charged leptons and neutrinos.
Other symmetries would play the same role in string models.

\begin{table}[h]
\begin{tabular}{|c|cc||ccccc|}
\hline
&$(\ell_e,\ell_\mu,\ell_\tau)$ & $(e^c,\mu^c,\tau^c)$  & $H_{u,d}$  
&$\chi_1$&$\chi_2$
&$(\chi_3,\chi_4,\chi_5)$&$(\chi_6,\chi_7,\chi_8)$   \\ \hline
$\Delta(54)$ & $3$ & $3$  & $1$& $1$& $1$&$3$&  $3$   \\
$Z_{3}$ & $1$ & $0$  & $0$&$2$
&$1$&$2$&  $1$   \\
\hline
\end{tabular}
\caption{Matter content and charge assignment of the $\Delta(54)$ model.}
\end{table}

The tensor products of $\Delta(54)$ triplets are the same as 
those of $S_4$.
Then, we realize the same superpotential (\ref{eq:model4-W-L}) 
and (\ref{eq:model4-W-nu}).
Thus, we can obtain the same results as one in section 4 
when the same vacuum alignment is realized.

The $\Delta(54)$ triplet corresponds to localized fields on 
three fixed points of the $Z_3$ orbifold in heterotic models.
Thus, the three families of left and right-handed leptons 
as well as triplet flavons, $(\chi_3,\chi_4,\chi_5)$ and 
$(\chi_6,\chi_7,\chi_8)$, would correspond to 
the modes localized on the three $Z_3$ fixed points.
Since the $\Delta(54)$ trivial singlet corresponds to 
a bulk mode on the orbifold, 
the electroweak Higgs fields and singlet flavons, $\chi_1$ and
$\chi_2$, are originated from the bulk modes.
Furthermore, VEVs of scalar fields on a fixed point 
correspond to blow-up of the orbifold singularity.
That is, our model suggests that a certain type of 
blow-up from the orbifold limit to Calibi-Yau manifold 
would be interesting to derive realistic lepton mass matrices, 
such that the flavon VEVs corresponding to 
(\ref{eq:model-4-vacuum}) are realized.
Thus, our model would be useful for model building from 
string models, too.

\section{Conclusion}

We have studied $S_4$ models with smaller numbers of flavon 
fields and free parameters.
When we introduce one $S_4$ triplet flavon, 
a realistic model cannot be constructed. 
To be consistent with experiments, 
we need two triplet $S_4$ flavons, or one triplet and one doublet 
at least. 
By building models with two triplets, or one triplet and 
one doublet, we have stronger predictions among 
lepton masses and mixing angles.

Realistic and predictive models are model III and model IV. 
Both models have seven parameters among 
six lepton mass eigenvalues and three mixing angles.
We have assumed there is a vanishing CP-phase in the lepton sector 
to make stronger prediction. However, it is easy to 
extend the models with non-vanishing CP-violation.

We can construct the model with the $\Delta(54)$ flavor symmetry, 
which is quite similar to model IV.
Such a model is quite interesting from the viewpoint of 
stringy realization.
We would study elsewhere on this aspect.

\vskip 1cm
{\bf Note to be added}

After this paper was completed, Ref.~\cite{:2011sj} appeared 
showing the range of the mixing angle $\theta_{13}$ in 
the latest T2K experiment.
Our prediction of $\theta_{13}$ is compatible with their result.

\vspace{1cm}
\noindent
{\bf Acknowledgement}

H.I. is supported by Grand-in-Aid for Scientific Research,
No.23.696 from the Japan Society of Promotion of Science. 
T. K. is supported in part by the Grant-in-Aid for Scientific 
Research No. 20540266 and the
Grant-in-Aid for the Global COE Program ``The Next 
Generation of Physics, Spun from
Universality and Emergence'' from the Ministry of Education, 
Culture,Sports, Science and Technology of Japan.

\end{document}